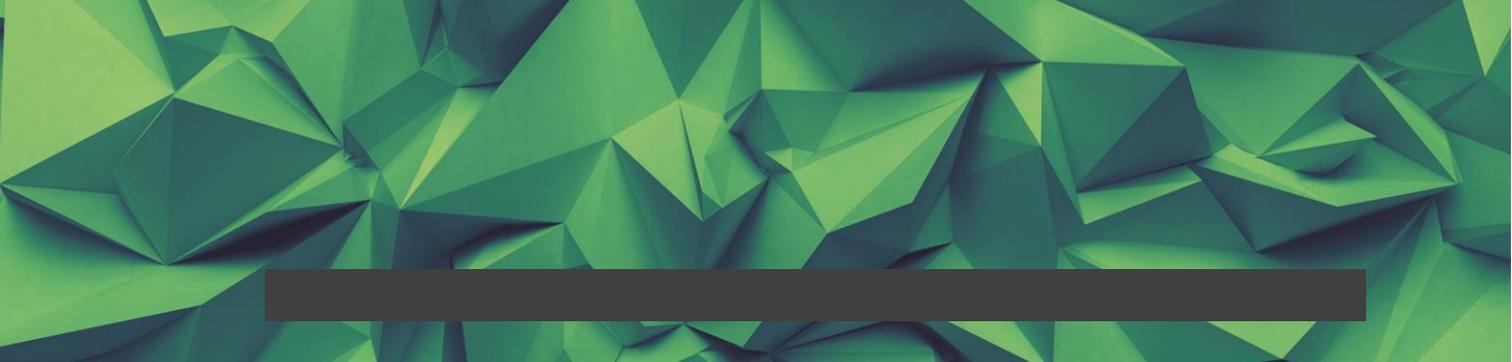

# User-centric Privacy Engineering for the Internet of Things

**Mahmoud Barhamgi**
Claude Bernard University

**Charith Perera**
Cardiff University

**Chirine Ghedira**
Lyon 3 University

**Djamal Benslimane**
Claude Bernard University

User privacy concerns are widely regarded as a key obstacle to the success of modern smart cyber-physical systems. In this paper, we analyse, through an example, some of the requirements that future data collection architectures of these systems should implement to provide effective privacy protection for users. Then, we give an example of how these requirements can be implemented in a smart home scenario. Our example architecture allows the user to balance the privacy risks with the potential benefits and take a practical decision determining the extent of the sharing. Based on this example architecture, we identify a number of challenges that must be addressed by future data processing systems in order to achieve effective privacy management for smart cyber-physical systems.

## 1. INTRODUCTION

We progressively find ourselves surrounded by smart cyber-physical systems that silently track our activities and collect sensitive information about us. Among the most prominent examples, we cite smart energy grids, smart transportation networks, smart homes and cities, etc. However, while such systems promise to ease our lives, they raise major privacy concerns for their users, as the data collected is often privacy-sensitive, such as location and habits of individuals, patients' vital signs, etc. In fact, collected data could be misused by the providers of such systems or even sold to interested third parties and exploited for various purposes [1, 2].

Recent privacy and data protection laws [3] called for more involvement of users in protecting their data by enabling them to control what is collected about them, when, by whom and for what purposes. However, existing solutions [4, 5, 6, 7] for privacy protection in cyber-physical systems fall short of that objective. Most of these solutions are inspired by the old approach employed in databases to ensure the privacy of users [8]. In that approach, the user is required by the system to provide her data, then she is prompted to specify her privacy preferences (through a set of variables) as to who can access the data and for what purposes, and to accept a privacy policy specifying a set of rules that may refer to her preferences. Subsequently, the rules are applied to all queries received by the system before returning the result. Unfortunately, such an approach does not provide effective protection of privacy. In fact, the user does not always understand the privacy policy [9], which could be incomplete, and may not be necessarily aware of the direct and indirect risks that may be associated with the disclosure of her data to a given entity in order to correctly specify her privacy preferences.

The successful involvement of users in protecting their data and ensuring their privacy requires two main conditions to be met. First, the user should be empowered to understand the privacy risks associated with the disclosure of a piece of data to a given entity and to balance these risks with the potential benefits of the disclosure to be able to take a meaningful privacy decision as to disclose or not, and to what extent. Privacy decisions are intrinsically difficult due to their delayed and uncertain consequences that are hard to compare with the immediate rewards of data disclosure. Second, the user should be provided with the necessary tools to implement her privacy decision by controlling the disclosure level. For example, different data degradation strategies may be used to modify the accuracy of the to-be-disclosed data item to achieve the chosen trade-off between the risks and the benefits.

In this article, we present a vision of how users of smart cyber-physical systems (e.g., Internet of Things) can be empowered to take a central and effective role in protecting their privacy. We materialise our vision by the proposition of a reference data sharing architecture that allows users to take flexible and practical data sharing decisions that reconcile their privacy requirements with their desires to be rewarded. We also validated our vision by implementing the proposed architecture within a smart environment to monitor chronic patients at home.

Compared to similar works [4,5,6,7] that sought to involve users in protecting their data such our solution has the following advantages. First, it allows end-users to assess the implicit privacy risks that are associated with the release of their data, compare them in a meaningful way with the benefits to choose the best data protection level. Second, privacy protection is ensured in a pragmatic way, allowing users to take a pragmatic stance between their interests and the risks implied. Third, our solution does not protect privacy in a rigid way, i.e., as context changes, inferred privacy risks change as well, and so is the protection ensured by the solution, allowing for more responsiveness to surrounding IoT environment.

The remainder of the paper is organized as follows. Section 2, analyses some of the key requirements for effective privacy protection in smart cyber-physical systems. Sections 3 and 4 present an example of an architecture implementing the identified requirements in a smart home scenario. Section 5 concludes the paper by pointing out future research directions.

# 2. PRIVACY FOR SMART CYBER-PHYSICAL SYSTEMS – A WALKING THROUGH EXAMPLE

Assume that Alice is the owner of a smart home featuring different types of smart appliances including: a refrigerator, stove, microwave, etc. Alice is a CDK (Chronic Kidney Disease) patient and has a home haemodialysis machine. The environment includes also smart meters for measuring the energy consumption, and different types of sensors including light, presence and temperature sensors, etc. These appliances, sensors and meters generate important data volumes that could be exploited by different entities for achieving different purposes.

*Utilities and edge services:* The electricity provider of Alice would be interested in exploiting the generated data (e.g., the energy consumption) to improve the energy distribution across the city and avoid service cut-offs. It can also use it to provide Alice with personalized recommendations for reducing her energy consumption and bills. Edge services include businesses providing services to energy consumers based on energy consumption data. Examples of services include, real-time energy usage monitoring to optimize the consumption (e.g., by proposing actions to users such as turning on/off a certain device, etc.), raising the energy awareness by allowing consumers to monitor their carbon emissions and compare them to those of friends on social media, etc.

*Law enforcement agencies*: They may use the collected data for different purposes, such as performing real-time (or near real-time) surveillance on suspects by determining if they are present and their current activities inside the home. Police investigators may also screen the energy consumption records of the utility to identify houses where some illegal activities may take place, e.g., potential drug production sites across the city.

*Marketers*: Marketers may be interested in determining the living profile of Alice to send her targeted advertisements. For example, they may be interested in knowing the appliances she may or may not own, or her eating patterns to send her special offers, etc.

## 2.1 Privacy concerns

The cited possible data uses may raise several privacy concerns for Alice. First, even though Alice may be interested in reaping the benefits offered by a utility or some edge services, like for instance the usage optimization of her appliances, either by explicitly providing detailed information about the appliances' usage or by providing energy consumption data that is granular enough to infer the appliances usage (through their consumption signatures [10]), she may not want to disclose the fact that she is a CKD patient (through the use of her haemodialysis machine). Such sensitive information, if was disclosed inadvertently or intentionally by one of the entities processing her data, would irreversibly affect her professional and social life. Second, Alice may not wish to disclose her habits and living style, e.g., presence/absence, walking/sleeping, eating and watching TV patterns, etc. Such information, if misused, would harm her in different ways, e.g., she might become a target for housebreakers, or get penalized by her employers and/or her social entourage, etc. Alice, may not wish to be under permanent surveillance. This would make her feel uncomfortable and impact her natural behaviour.

In this work, we uphold the privacy definition given in [11], where the privacy is described as consisting of four dimensions: (1) *privacy of the personal information*, (2) *privacy of the person* (i.e., the integrity of her body), (3) *privacy of personal behaviour* and (4) *privacy of personal communications*. We therefore consider all the data items that may be used to compromise the privacy of a person in at least one of these dimensions as sensitive.

## 2.2 Practical privacy decisions

Alice is also a pragmatic person and may be willing to trade off the anticipated benefits (of sharing some of her data with some of the cited entities) with the potential privacy risks and make compromises. For example, she may accept to release fine-grained energy consumption data (i.e., consumption readings with high sampling frequency) to her electricity provider to help it better optimize its energy distribution, provided some financial benefits (e.g., bill reductions or bonuses, etc.). She may also accept to share with an edge service only the data necessary to compute her daily carbon emission and compare it with those of friends on a social network (to gain some social recognition), but not to infer the list of owned appliances. Similarly, she may accept to provide law enforcement services with the necessary data to check for illegal activities, but not to perform real-time surveillance.

In all of these examples, Alice takes her practical decision after evaluating how trustworthy the data consumer is, and for what purposes (that potentially relate to her privacy requirements and preferences) the released data can be exploited, and what are the benefits of the data sharing decision.

## 2.3 Identified requirements

Based on our simple example, we identify below some of the key requirements for ensuring effective privacy protection for cyber-physical systems.

***R1: User-centric privacy protection:*** As required by data protection regulations, users (i.e., data owners) should be enabled to play a central role in protecting their data. This implies that before sharing a data item with a given data consumer, the user should be (*i*) enabled to understand the privacy risks that pertain to her subjective vision of privacy, and (*ii*) provided with the necessary mechanisms to control the extent of the sharing. For example, the data sharing architecture should warn Alice, before releasing her fine-grained energy consumption readings to her electricity provider or to an edge service, that these bodies may determine her appliances (based on their signatures) and consequently know that she is a CDK patient. It should also allow Alice to reduce the frequency of released energy consumption readings to prevent such privacy breach.

***R2: Pragmatic privacy decision making:*** Studies [12] showed that users, in practice, tend to take a pragmatic stance on sharing their private data, i.e., they would accept the release of some of their private data in return for some incentives. This has motivated an important drive in the database research field to monetize private data, based on potential privacy risks, and compensate end users directly [13]. Based on that observation, privacy protection in cyber-physical systems should not be ensured in a rigid fashion by deciding whether a data item should be shared with a given entity or not. Rather, users (e.g., Alice) should be empowered to assess the privacy risks that they accept to take, balance them against the benefits offered by data consumers, and potentially negotiate with data consumers before taking their sharing decisions. Data owners and consumers will play roles very much similar to the roles of sellers and buyers in a free market, where buyers and sellers may bargain with each other to reach a suitable deal.

***R3: Adaptive privacy protection to cope with context evolution:*** The data sharing decision may depend on user's context. For example, Alice may accept, in the general case, to release fine-grained energy consumption data to her electricity provider that would allow the latter to infer her appliances usage, but not when she uses her haemodialysis machine. The data sharing architecture should detect the context changes that would lead to privacy breaches and take the necessary actions to avoid them. For instance, the architecture could warn Alice when she turns on her medical machine about the potential leakage of her health condition to her energy provider and could propose to her to decrease the sampling frequency of the released energy consumption readings.

# 3. A REFERENCE DATA SHARING ARCHITECTURE FOR USER-CENTRIC PRIVACY PRESERVATION FOR IoT

In this section, we describe an example of a data sharing architecture for a smart home that satisfies the requirements discussed above and gives the control over data sharing back to end-users.

## 3.1 Architecture overview

We use the terms "*data owners*" to designate the users of cyber-physical systems that generate data by interacting with the systems (e.g., occupants of smart homes, monitored patients, etc.), and "*data consumers*" to designate the stakeholders that are interested in collecting and exploiting the data generated, such as electricity companies in smart grids, healthcare providers in intelligent healthcare networks, third-parties that would provide data owners with smart services, etc.

In our architecture (Figure-1) raw data generated by connected things (IoT objects) is stored within a *Personal Data Store* before released to data consumers. All data flows between data owners and consumers pass by the proposed reference architecture. When a data consumer queries a data item from a data owner (either directly, or by getting the data owner to use a service/application provided by data consumer), our architecture processes the received query as follows:

- It assesses, through the *Privacy Risks Inference component* (or the *Privacy Oracle*), the risks associated with releasing the requested data to data consumer. Since privacy is a subjective

- notion (i.e., different people may be concerned with different privacy risks), the privacy oracle takes into account several factors including the profile of data owner, her context, her trust in data consumer, etc. Context is a key input to the privacy oracle, it is monitored in a continuous way (by observing the data sent by IoT objects) and exploited in identifying the risks.
- Then, it *monetizes* (i.e., quantifies) the identified privacy risks and the potential benefits using a numerical model, through the *Trade-off Data Sharing component (TDS)*, and helps data owner take a pragmatic decision balancing the two. The decision denotes the data items that can be shared with the consumer, along with their accuracy (i.e., precision). Data owners and consumers could also negotiate before reaching a pragmatic data sharing decision.
- Finally, it modifies, through the *Query/Result Modification component (QRM)*, the query before being applied to the personal data store to discard the data items to which the data consumer is not entitled, and the query's result to change its accuracy before its release.

Privacy risks closely relate to what can be inferred from collected data. For example, granular readings of smart electricity meters can be analysed to infer information about the occupants such as their presence/absence, the possession of specific devices. The disclosure of such information could lead to privacy risks such as being subject to discrimination, surveillance, burglaries, etc. Our architecture identifies the relevant risks, *through the Privacy Risks Inference* component, by rendering the implicit relationship between raw collected data and risks explicit.

Context changes could lead to privacy breaches in previously taken decisions. Therefore, context is monitored and risks are analysed in a permanent way (i.e., old data sharing decisions might be recomputed when context changes).

Our architecture would use any data protecting scheme (whose efficacy is proved) to anonymise the data once a data sharing decision is taken. For examples, differential privacy can be used to alter the precision of released smart meter readings, anonymization can be used to anonymise the location, etc.

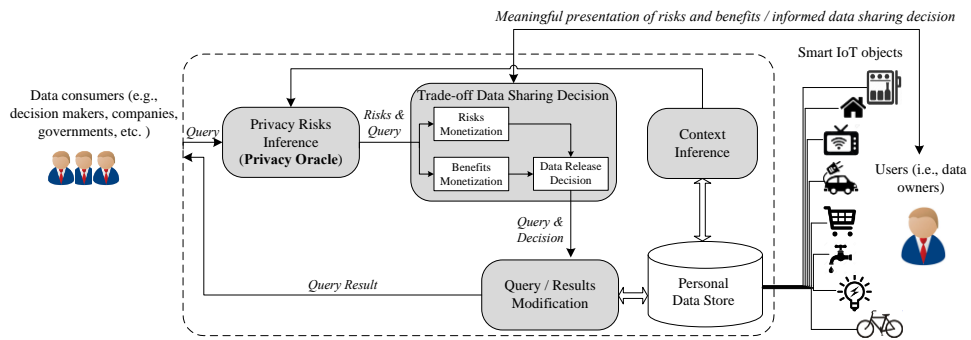

**Figure 1: A reference architecture for allowing the users of IoT smart systems to control their data**

## 3.2 Privacy risks/benefits trade-off model

We present in this subsection our model to trade off the privacy risks with the data sharing benefits. Our model builds on previous works on trade-off decision models such as [14]. The model is shown in Figure 2. The trade-off decision is taken based on two factors: the *privacy risks* associated with answering the query $q$ of the data consumer $d$, and the *benefits* generated by the query answering. The *privacy risks factor* has a negative impact on the trade-off decision, whereas the *benefits factor* has a positive one. The *privacy risks* are measured based on three factors: (1) the *sensitivity of the data items* requested in $q$, (2) the *trust in the recipient* $d$ (i.e., data consumer), and (3) the *information leakage* caused by answering $q$. We will define all of these factors, and how they can be computed in subsequent subsections. We define below the trade-off privacy decision.

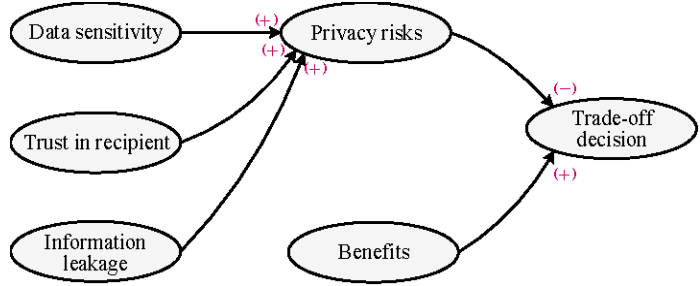

Figure 2: Privacy risks/benefit trade-off model

*Definition 1:* (**Trade-off privacy decision**) the decision to answer the query $q$ of the data consumer $d$ is computed as follows:

$$Decision\ (q, d) = \begin{cases} \boldsymbol{Answer}, & if\ U_d(q) > 0 \\ \boldsymbol{Deny}, & otherwise \end{cases}$$

where $U_d(q)$ is the utility function of answering $q$, and it is computed as follows:

$$U_d = (1 - w) * b_d(q) + w * r_d(q)$$

where $b_d(q)$ and $r_d(q)$ are functions quantifying the benefits and the risks that are generated/caused by answering $q$, respectively. The parameter $w$ can be tuned by the user to bias the benefits/risks trade-off decision (e.g., if $w = 0.5$, the benefits should be at least equal to the risks). In the following we detail how $r_d(q)$ and $b_d(q)$ can be computed.

### 3.2.1 Measuring the privacy risks

The privacy risks can be measured based on three factors: the sensitivity of queried data items, the trust in the data recipient and the information leakage caused by the release of the data items. We discuss these factors in the following.

*Data sensitivity*: the sensitivity of a data item can be determined based on its direct or indirect relations with a privacy sensitive information. For example, the electricity consumption readings are not privacy sensitive in themselves, but the fact that they can be analysed by some data mining algorithms [10] to determine the appliances usage and consequently infer the behaviour patterns (e.g., waking/sleeping patterns, meal times, etc.) make them sensitive.

The NIST guidelines for smart grid cybersecurity have identified the different privacy-sensitive information pieces that could be relevant to a wide range of users in smart home scenarios and that correspond to our vision of privacy. These include: *user personal information*, *presence/absence*, *real-time surveillance*, *habits*, and *the use of a specific device*. We call these information pieces as the *privacy parameters* and exploit them to compute the sensitivity of a data item $i$ (e.g., energy consumption data, etc.) as follows. The user is prompted to assign an importance weight to each of the privacy parameters $f_j (1 \leq n)$. Then, whenever a data item $i$ is requested by a query, its sensitivity is computed by summing up the weights of all the privacy parameters that relate to (i.e., can be inferred from) it by the equation:

$$\text{Sensitivity (i)} = \sum_{j=1}^{n} weight\ (f_j) * context\ (f_j)$$

where the function **context($f_j$)** $\in$ *{0,1}*, it takes the value "1" or "0" depending on whether $f_j$ is relevant or not in the current context.

*Trust in recipient*: the trust of a user in a data consumer may change depending on the consumer's profile and its history of interaction with users. For example, a law enforcement agency may be trusted more than a publicity company, and an electricity provider with which a user has had a long and satisfactory history of interaction may be trusted more than an unknown edge service. In our model, we can rely on any of the trust models (e.g., [15]) that compute the trust in a consumer by aggregating and averaging quantitative feedback ratings of users (e.g., smart house owners).

*Information leakage*: it measures the knowledge leaked to $d$ about the different privacy sensitive parameters $f_j$ ($1 \leq n$) when a data item $i$ is disclosed to $d$. The information leakage relative to a particular privacy parameter $f_j$, denoted by $L_{f_j}$, can be measured by capturing the uncertainty of $d$ about $f_j$ and is dependent on the accuracy level of the released $i$. $L_{f_j}$ can be measured either analytically or experimentally. For example, the experimental studies in [10] showed that releasing the energy consumption with a good accuracy (i.e., sampling frequency of 15 seconds) leads to determining the eating habits with a confidence degree of 59% and the appliance use with a confidence degree of 72%, but when the sampling frequency drops to 30 minutes these two confidence degrees drop to near 0.

### 3.2.1 Measuring the benefits

Different types of benefits require different kinds of measurements. For example, the usage optimization advices that would lead to less energy consumption, bonuses and bill reductions that one could receive from her electricity provider can be quantified by their introduced financial gain. The ben-efits of sharing fine-grained data with law enforcement agencies to contribute to securing the living environment could be quantified by the one's feeling of self-satisfaction and patriotism. The benefits of sharing her carbon emission on a social network can be quantified by the social recognition she receives.

# 4. REAL-LIFE CASE-STUDY: MONITORING ELDERLY AND CHRONIC PATIENTS

We implemented our reference architecture within a smart environment for monitoring chronic patients. The environment involves wearable sensors for monitoring several vital signs such as heart rates, blood pressure, ECG as well as smart objects and sensors installed in fixed positions of the monitored environment. The study involved 20 real patients with ages from 20 to 67 years.

Figure 3 shows the implementation architecture. Collected raw data are stored within a personal data vault (PDV) that provides a simple implementation of the different component of our architecture. Healthcare providers provide patients, through a dedicated mobile application, with personalized healthcare services by consuming collected data. Data flows between users (i.e., patients) and data consumers (healthcare providers) go through the PDV.

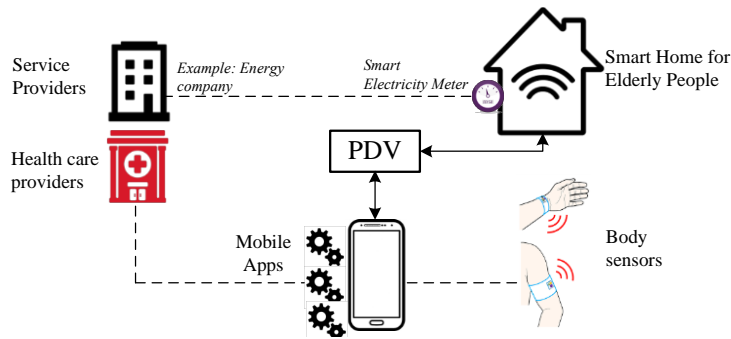

Figure 3: Implementation of the proposed architecture within an IoT environment for monitoring chronic patients

We implemented the privacy risk inference component as a knowledgebase. Collected data, context elements and associated risks are modelled by a domain ontology. The knowledgebase models the implicit relationship between raw data and risks by a set of inference rules (expressed in the Semantic Web Rule Language SWRL). We present in Figure 4 simplified examples of inference rules. Rule-1 states that the use of a device can be inferred from the readings of an energy smart meter (EMR). The terms "*Person*", "*EMR*", "*Device*" are ontological concepts, whereas "*hasEMR*", "*isShared*", "*useDevice*", "*isInferrable*" are properties. Rule-2 states that the use of a medical device reveals the health conditions for which the device is used. Rule-1 and Rule-2 can be combined together to infer the fact that the health conditions could be inferred from the readings of an energy smart meter. Rule-3 simply states that "*location*" and "*heart-rate*" metadata could be combined to infer if data owner has an extramarital affair (i.e. heart-rate can be used to infer if data owner is engaged in a sexual activity, and location can be exploited to infer whether data owner is in a suspicious location (e.g., outside home)). The rule uses contextual information such as whether the data owner is married.

*Rule-1 (Inferring a device usage by Energy Meter Readings EMR):*

*Person(?p), hasEMR(?p,?r), EMR(?r), isShared(?r, "true")*
➡ *useDevice(?p, ?d), Device(?d), isInferrable(?d,"true")*

*Rule-2 (Inferring health conditions by a Medical device):*

*Person(?p), usesDevice(?p,?r), MedicalDevice(?r), cures(?r,?m), disease(?m)*
➡ *hasDisease(?p, ?m)*

*Rule-3 (Inferring extramarital affairs by location and sexual activity):*

*Person(?p), isMarried(?p, "true") hasASexualActivity (?p,?x), isInferable(?x, "true"), hasLocation(?p,?l), isSuspicious(?l, "true")*
➡ *has(?p, ?f), ExtramaritalAffair(f), isInferable(?f, "true")*

Figure 4: Sample of inference rules for inferring privacy risks.

Figure 5 (Window-1) shows the user interface of the PDV. Upon the reception of a new data request, the PDV takes into account the user's context and her shared data (Window-2) to provide her with a description of associated privacy risks (Window-3) as well as a set of recommended actions (Window-4).

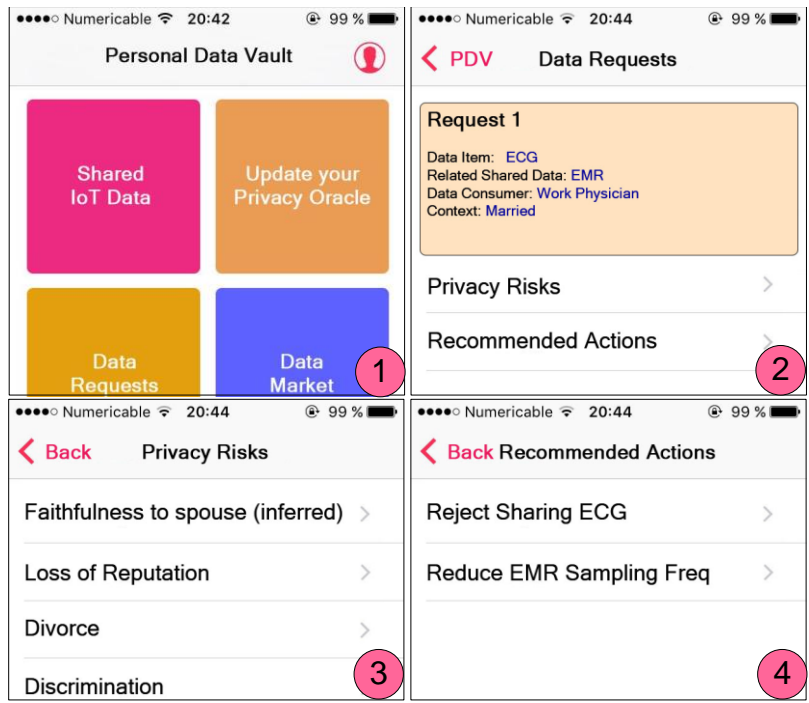

Figure 5: Interaction with end-users to control the release of data

## 5. CHALLENGES AND OPEN ISSUES

We conclude the paper by identifying some of the key challenges that should be addressed to achieve effective privacy protection in cyber-physical systems.

*Meaningful data degradation strategies*. Smart environments produce a rich set of data elements that are different in nature and require different forms of degradation (to implement chosen trade-off decisions). For example, data elements such as the *address* or the *age* of a person can be degraded by applying well known data anonymization techniques, whereas the *electricity consumption readings* can be degraded by reducing the sampling frequency. Research efforts are needed to define for each type of data elements suitable degradation strategies, and data degradation levels that map directly to their different possible uses (i.e. risks), and to data leakage levels.

*Rich pricing models for privacy sensitive data*. Data consumers may offer different forms of benefits including financial, social, societal benefits, etc. Research efforts are needed to devise rich and flexible pricing models (for privacy sensitive data) that would fit for various forms of benefits. Efforts are also needed to help users assess the sensitivity of their private data. Crowdsourcing techniques can be explored for that purpose.

*Context modeling and monitoring for triggering the adaptation*. The sensitivity of a data piece may depend on the context. For example, the energy consumption data may become sensitive when Alice turns on her hemodialysis machine. Models and techniques are needed to represent and monitor the context elements that relate to user's privacy and detect context changes that require adapting privacy decisions. Complex Event Processing CEP techniques can be explored for that purpose. In fact, the interaction of users with their surrounding environment generates various events that can be monitored. The contexts that require adaptation are represented by a set of events combinations that can be tracked by the CEP system.

*Mahmoud Barhamgi* is an associate professor at Lyon-1 University (France) whose research focuses on Privacy preservation in SOA, Web and Cloud environments. Mahmoud.barhamgi@univ-lyon1.fr

*Charith Perera* is a lecturer at Cardiff University (UK) whose research focuses on Security and Privacy in IoT. charith.perera@ieee.org

*Chirine Ghedira* is a professor at Lyon-1 University (France) whose research focuses on Service-oriented architectures. chirine.ghedira-guegan@univ-lyon3.fr

*Djamal Bensliamne* is a professor at Lyon-1 University (France) whose research focuses on Service-oriented architectures. Djamal.benslimane@univ-lyon1.fr